\documentstyle[sprocl]{article}
\begin{document}

\newcommand{\ra}{\rightarrow}
\newcommand{\ko}{K^0}
\newcommand{\be}{\begin{equation}}
\newcommand{\ee}{\end{equation}}
\newcommand{\bea}{\begin{eqnarray}}
\newcommand{\eea}{\end{eqnarray}}

\title{$p$--BRANES, $D$--BRANES AND $M$--BRANES}

\author{ E.~BERGSHOEFF }

\address{Institute for Theoretical Physics, University of Groningen,\\
Nijenborgh 4, 9747 AG Groningen, The Netherlands}

\maketitle
\abstracts{
We consider solutions to the string effective action corresponding to
$p$--Branes, $D$--Branes and $M$--Branes and discuss some
of their properties.}

\section{Introduction}

The original classification of super $p$--branes was based on the assumption
that the embedding coordinates, in a physical gauge, form worldvolume
scalar multiplets \cite{Ac1}. A classification of such scalar multiplets 
with $T$ (transverse) scalar degrees of freedom in $p+1$
dimensions leads to the following table:

\begin{table}[h]
\caption{Scalar multiplets with $T$ scalars in 
$p+1$ dimensions.\label{tab:sc}}
\vspace{0.4cm}
\begin{center}
\begin{tabular}{|c|c|c|c|c|}
\hline
$p+1$&$T$&$T$&$T$&$T$
\\ \hline
1&1&2&4&8\\
2&1&2&4&8\\
3&1&2&4&8\\
4&&2&4&\\
5&&&4&\\
6&&&4&\\ 
\hline
\end{tabular}
\end{center}
\end{table}

\noindent Each scalar multiplet corresponds to a $p$--brane in $d$ target
spacetime dimensions with 

\begin{equation}
d= (p+1) + T\, .
\end{equation}
Here the target space has been divided into
$p+1$ worldvolume and $T$ transverse directions.
The Table describes 16 $p$--branes with $p\le 5$ and $d\le 11$.

The corresponding $p$--brane actions consist of a kinetic term and
a Wess-Zumino (WZ) term. The kinetic term is given by

\begin{equation}
S_{\rm kin}^{(p)} = \int d^{p+1} \xi \ \sqrt {|g|}\, ,
\end{equation}
where $g$ is the determinant of the induced metric. The WZ term
is given by the pull-back of a $(p+1)$--form potential.
For the
case of interest (10 dimensions) these are a two--index
Neveu-Schwarz/Neveu-Schwarz (NS/NS) tensor $B^{(1)}$ (heterotic,
IIA or IIB one--brane) 
and the dual six-index tensor ${\tilde B}^{(1)}_{\rm het}$ (heterotic
five--brane):

\begin{eqnarray}
S_{\rm WZ}^{(1)} &=& \int d^{2}\xi\  B^{(1)}\, ,\label{WZ1}\\
S_{\rm WZ}^{(5)} &=&  \int d^{6}\xi\ {\tilde B}^{(1)}_{\rm het}\, ,
\label{WZ5}
\end{eqnarray}
where $d {\tilde B}_{\rm het}^{(1)} = {}^* d B^{(1)}$. Note that the action 
(\ref{WZ1}) is the same independent of whether the one--brane is
propagating in a $N=1$, IIA or IIB supergravity background.

\section{Solutions }

To each $p$--Brane given in Table 1 one can associate 
a solution corresponding to the following
$d$--dimensional string--frame Lagrangian:

\begin{equation}
\label{toy}
{\cal L}_{S,d} = \sqrt {|g|}\biggl \{e^{-2\phi}\bigl [
R -4(\partial\phi)^2\bigr ] + {(-)^{p+1}\over 2(p+2)!}
e^{a\phi}F^2_{(p+2)}\biggr\}
\, .
\end{equation}
Here $\phi$ is the dilaton and $F_{(p+2)}$ is the curvature of a
$(p+1)$--form gauge field
\begin{equation}
F_{(p+2)} = d A_{(p+1)}\, .
\end{equation}
Note that we use a form of the action
in which both the (electric) $(p+1)$-form gauge field as well as
the (magnetic) $(D-p-3)$-form gauge field occurs. It is understood
that the duality relation between them is only applied at the
level of the equations of motion and not in the action. This is
similar to the pseudo-action formulation of \cite{Be1}.

The general $p$--Brane solution is given by \cite{Gi1}

\begin{eqnarray}
\label{sol1}
ds_{S,d}^2 &=& H^\alpha dx^2_{(p+1)} - H^\beta dx^2_{(D-p-1)}\, ,\nonumber\\
e^{2\phi} &=& H^\gamma\, ,\\
F_{0\cdots pi} &=& \delta\ \partial_i H^\epsilon\, ,\nonumber
\end{eqnarray}
with the parameters $\alpha, \cdots ,\epsilon$ given by

\begin{eqnarray}
\label{sol2}
\alpha &=& {1\over N} (2-a)\, ,\hskip .6truecm 
\beta = -{1\over N} (2+a)\, , \nonumber\\
\gamma &=& {1\over N} \bigl [2(p+1) +(2+a)(1-{1\over 2}d)\bigr ]\, ,\\
\delta^2 &=& -{4\over N}\, , \hskip 1.5truecm \epsilon = -1\, ,\nonumber
\end{eqnarray}
with $N = (p+1)a + (1-{1\over 2}d)(1+{1\over 2}a)^2$.
In the case of $p$--branes it is assumed that the $(p+1)$--form
gauge field is a NS/NS field, i.e. $a=-2\ (+2)$ for the 
electric (magnetic) gauge-fields. For instance, the 
ten--dimensional one--brane, or fundamental string, solution is given 
by (take $d=10, p=1$ and $a=-2$) \cite{Da1}

\begin{eqnarray}
\label{sol3}
ds_{S,10}^2 &=& H^{-1} dx^2_{(2)} - dx^2_{(8)}\, ,\nonumber\\
e^{2\phi} &=& H^{-1}\, ,\\
F_{01i} &=& \partial_i H^{-1}\, ,\nonumber
\end{eqnarray}
where $F_{(3)}$ is the curvature of the NS/NS tensor $B^{(1)}$.
On the other hand, the ten--dimensional five--brane solution corresponds to
(take $d=10, p=5$ and $a=+2$)

\begin{eqnarray}
\label{sol4}
ds_{S,10}^2 &=& dx^2_{(6)} - H dx^2_{(4)}\, ,\nonumber\\
e^{2\phi} &=& H\, ,\\
F_{012345i} &=& \partial_i H^{-1}\, .\nonumber
\end{eqnarray}
Note that in the latter case we use the dual six-index tensor
$\tilde {B}_{\rm het}^{(1)}$.

Another example is provided by $p$--Brane solutions in eleven dimensions
which we refer to as $Mp$--Branes \cite{To1}.
Since there is no dilaton in this case, it is more convenient to
work with the Einstein--frame Lagrangian

\begin{equation}
{\cal L}_{E,d} = \sqrt {|g|}\biggl [ R +{1\over 2}(\partial\phi)^2
+ {(-)^{p+1}\over 2(p+2)!}e^{a\phi}F^2_{(p+2)}\biggr ]\, ,
\end{equation}
in which case the general $p$--Brane solution is given by

\begin{eqnarray}
\label{solE2}
\alpha &=& -{4\over N}(d-p-3)\, ,\hskip 1.5truecm
\beta = {4\over N} (p+1)\, ,\nonumber\\
\gamma &=& {4a\over N} (d-2)\, ,\hskip .5truecm
\delta^2 = {4\over N} (d-2)\, , \hskip .5truecm
\epsilon = -1\, ,
\end{eqnarray}
with $N = (d-2)a^2 + 2(p+1)(d-p-3)$.
For instance, the $M2$--Brane solution is given by (take $d=11, p=2$ and
$a=0$) \cite{St1}

\begin{eqnarray}
\label{sol5}
ds_{E,11}^2 &=& H^{-2/3} dx^2_{(3)} - H^{1/3} dx^2_{(8)}\, ,\nonumber\\
F_{012i} &=& \partial_i H^{-1}\, .
\end{eqnarray}

Recently, two different extensions of the $p$--Branes classified in 
Table 1 have been given. The first extension concerns the general
$Mp$--Branes in eleven dimensions. At the level of solutions,
it is clear that there exists also a $M5$--Brane solution given by
(take d=11, p=5 and $a=0$) \cite{Gu1}

\begin{eqnarray}
\label{sol6}
ds_{E,11}^2 &=& H^{-1/3} dx^2_{(6)} - H^{2/3} dx^2_{(5)}\, ,\nonumber\\
F_{012345i} &=& \partial_i H^{-1}\, .
\end{eqnarray}

The second extension concerns the so-called 
$Dp$--Branes \cite{Po1}. At the level of solutions they correspond to
$p$--Brane solutions whose charged is carried by a Ramond-Ramond (RR)
gauge field, i.e.~one which has $a=0$ in the string frame. The
$Dp$--solutions are most easily formulated as solutions of the
string-frame Lagrangian (\ref{toy}). These solutions are given by
(take $d=10, a=0$)

\begin{eqnarray}
\label{sol7}
ds_{S,10}^2 &=& H^{-1/2}dx^2_{(p+1)} - H^{+1/2} dx^2_{(9-p)}\, ,\nonumber\\
e^{2\phi} &=& H^{-{1\over 2}(p-3)}\, ,\\
F_{0\cdots pi} &=& \partial_i H^{-1}\, .\nonumber
\end{eqnarray}

\section{Actions}

The natural question arises why the $M5$-brane and all the $Dp$--branes
were absent in the original classification summarized in Table
1. As far as the $Dp$--Branes are concerned the answer is that these
extended objects 
are described by embedding coordinates that, in a
physical gauge, form worldvolume vector multiplets.
A classification of all vector multiplets with $T$ (transverse)
scalars in $p+1$
dimensions is given by the table below.

\begin{table}[h]
\caption{Vector multiplets with $T$ scalar degrees of freedom in 
$p+1$ dimensions.\label{tab:ve}}
\vspace{0.4cm}
\begin{center}
\begin{tabular}{|c|c|c|c|c|}
\hline
$p+1$&$T$&$T$&$T$&$T$
\\ \hline
1&2&3&5&9\\
2&1&2&4&8\\
3&0&1&3&7\\
4&&0&2&6\\
5&&&1&5\\
6&&&0&4\\ 
7&&&&3\\
8&&&&2\\
9&&&&1\\
10&&&&0\\
\hline
\end{tabular}
\end{center}
\end{table}
In the case of ten dimensions, this leads to $Dp$--branes for
$0\le p \le 9$. The kinetic term of these $Dp$--branes is given 
by the following Born--Infeld type action:

\begin{equation}
S_{\rm kin}^{(Dp)} = \int d^{p+1}\xi\ e^{-\phi} \sqrt {|{\rm det}
(g_{ij} + {\cal F}_{ij})|}\, ,
\end{equation}
where $g_{ij}$ is the embedding metric and ${\cal F} = 2 d V -
B^{(1)}$ is the curvature of the worldvolume gauge field $V$.  

There is also a WZ term which describes the coupling of the 
RR fields to the $Dp$--brane:

\begin{equation}
S_{\rm WZ}^{(Dp)} = \int d^{p+1}\xi \ {\cal A}\ e^{\cal F}\, ,
\end{equation}
with ${\cal A} = \sum_{q=0}^9 A_{(q+1)}$.
Here it is understood that, after expansion of the exponent,
the $(p+1)$--form is picked out. In particular it means that
the WZ--terms for $p \ge 3$ contain both electric as well as
magnetic potentials \cite{Gr1}. To define the dual potential
we need to specify whether the curved background is given by
$N=1$ or IIA/IIB supergravity. For instance, the dual
potentials that couple to the heterotic 5--brane, the IIA five--brane 
(which follows from direct dimensional reduction of the 
eleven--dimensional five--brane) and the D5--brane 
are defined by, respectively,

\begin{eqnarray}
{}^* dB^{(1)} &=& d {\tilde B}_{\rm het}^{(1)}
\, ,\nonumber\\
{}^* dB^{(1)} &=&
d {\tilde B}_{{\rm IIA}}^{(1)} - {\textstyle\frac{105}{4}} C d C - 
7 A^{(1)} G({\tilde C})\, ,\\
{}^* dB^{(1)} &=&
d {\tilde B}_{{\rm IIB}}^{(1)} +
D d B^{(2)} - {\textstyle\frac{1}{4}}\epsilon^{kl}
B^{(2)}B^{(k)} d B^{(l)}\, .\nonumber
\end{eqnarray}
Here $G({\tilde C})$ is the curvature of the dual 5--form potential
${\tilde C}$ defined in \cite{Be2}.

Similarly, for the $M5$--Brane 
the embedding coordinates, in a physical gauge,
form a worldvolume tensor multiplet. This leads us to the table below.

\begin{table}[h]
\caption{Tensor multiplets with $T$ scalar degrees of freedom in 
$p+1$ dimensions.\label{tab:te}}
\vspace{0.4cm}
\begin{center}
\begin{tabular}{|c|c|c|}
\hline
$p+1$&$T$&$T$\\ 
\hline
6&1&5\\
\hline
\end{tabular}
\end{center}
\end{table}
\noindent
Note that the $M5$--Brane in 7 dimensions has a one--dimensional transverse
space and hence is not asymptotically flat. A non--trivial feature of these
five--branes is that the tensor multiplet contains a selfdual two--form.
The kinetic term at quadratic order is given by \cite{To1}

\begin{equation}
S_{\rm kin}^{(M5)} = \int d^6\xi \sqrt {|g|} \biggl [ 1 + {1\over 2}
{\cal H}^2 + O({\cal H}^4) \biggr ]\, ,
\end{equation}
with  ${\cal H} = 3(dW-1/2C)$ and ${\cal H} = {}^*{\cal H} + O({\cal H}^3)$.
On the other hand, the WZ term is given by \cite{To1,Ah1,Be2}

\begin{equation}
S_{\rm WZ}^{(M5)} = \int d^6\xi \biggl [{\textstyle {1\over 70}}\tilde {C} +
{\textstyle {3\over 4}} {\cal H}C\biggr ]\, , 
\end{equation}
with the dual 6--form potential defined by \cite{Ca1}

\begin{equation}
d\tilde {C} - {\textstyle {105\over 4}} CdC = {}^* dC\, .
\end{equation}
It has been verified that, to quadratic order in ${\cal H}$, 
the $M5$--brane action
reduces to the $D4$--brane action \cite{Be2}. The determination of
the higher order in ${\cal H}$ terms of the kinetic term remains 
an open issue (see, however, \cite{Howe}).

\section{Supersymmetry}

It is not difficult to verify that all $p$-,$D$- and $M$--brane solutions
preserve half of the supersymmetry. Here we present a proof
for the $Dp$--branes. In order to give a unified treatment it is 
convenient to treat the IIA and IIB supergravity theories at an
equal footing (for more details, see \cite{Be3}).
The relevant supersymmetry rules of the gravitino and dilatino
\cite{Be4} are given by (using the string-frame metric)

\begin{eqnarray}
\delta \psi_\mu &=& \partial_\mu\epsilon - 
{1\over 4}\omega_\mu{}^{ab}\gamma_{ab}\epsilon
+ {(-)^p\over 8(p+2)!}e^\phi F\cdot \gamma\
\gamma_\mu\epsilon_{(p)}^\prime\, ,\nonumber\\
\delta\lambda &=& \gamma^\mu\bigl (\partial_\mu\phi \bigr )\epsilon +
{3-p\over 4(p+2)!} e^\phi F\cdot \gamma\ \epsilon_{(p)}^\prime\, ,
\end{eqnarray}
\bigskip
with $F\cdot \gamma \equiv F_{\mu_1\cdots \mu_{p+2}}\gamma^{\mu_1
\cdots \mu_{p+2}}$ and $\epsilon_{(p)}^\prime$ given by

\begin{table}[h]
\caption{Definition of the spinor $\epsilon_{(p)}^\prime$.\label{tab:tensor}}
\vspace{0.4cm}
\begin{center}
\begin{tabular}{|c|c|c|c|}
\hline
$p$&$\epsilon_{(p)}^\prime$\ (IIA) & $p$&$\epsilon_{(p)}^\prime$\ (IIB)\\
\hline
$0$ &$\epsilon$&$-1$&i$\epsilon$\\
$2$&$\gamma_{11}\epsilon$&$1$&$i\epsilon^\star$\\
$4$&$\epsilon$&$3$&i$\epsilon$\\
$6$&$\gamma_{11}\epsilon$&$5$&$i\epsilon^\star$\\
$8$&$\epsilon$&$ 7$&$i\epsilon$\\
 \hline
 \end{tabular}
 \end{center}
\end{table}
\bigskip

\noindent
A straightforward calculation shows that the Killing spinor is given by

\begin{equation}
\epsilon = H^{-1/8}\epsilon_0\, ,\hskip 1truecm 
\epsilon + \gamma_{01\cdots p}\ \epsilon_{(p)}^\prime = 0\, ,
\end{equation}
for constant spinor $\epsilon_0$.

\section{Open Issues}

Sofar we have encountered actions with worldvolume scalar, vector
and tensor multiplets. Consider the string and five-brane actions in 
ten dimensions. For each $p$ we expect a heterotic, IIA and two IIB
actions. The table below shows that, restricting ourselves to worldvolume
$(q+1)$-form gauge fields with $q=-1,0,1$\footnote{In the case of the
IIA five-brane there is a scalar that can be dualized to a 4-form.}, 
we cannot describe the
second IIB five-brane action. This action should, via $SL(2,R)$
duality, be related to the $D5$--brane action. A preliminary study \cite{Be5}
shows that the action involves a worldvolume 3--form gauge field,
as indicated in the table. It would be of interest to construct this action.

\begin{table}[h]
\caption{String and Five--brane actions with world-volume $(q+1)$--form 
gauge fields.}
\vspace{0.4cm}
\begin{center}
\begin{tabular}{|c|c|c|}
\hline
$q$&$p=1$&$p=5$\\ 
\hline
-1&het, IIA, IIB&het\\
1&IIB&IIB\\
2,4&&IIA\\
3&&IIB ?\\
\hline
\end{tabular}
\end{center}
\end{table}

\section*{Acknowledgments}

This work is supported by the European Commission TMR programme 
ERBF\hfill\break
MRX-CT96-0045, in which E.B. is associated to the University of
Utrecht. I thank J.P.~van der Schaar for providing me with the
string-frame analogue eq.~(\ref{sol2}) of eq.~(\ref{solE2}).

\section*{References}
\newcommand{\Journal}[4]{{#1} {\bf #2}, #3 (#4)}
\newcommand{\NCA}{\em Nuovo Cimento}
\newcommand{\NIM}{\em Nucl. Instrum. Methods}
\newcommand{\NIMA}{{\em Nucl. Instrum. Methods} A}
\newcommand{\NPB}{{\em Nucl. Phys.} B}
\newcommand{\PLB}{{\em Phys. Lett.}  B}
\newcommand{\PRL}{\em Phys. Rev. Lett.}
\newcommand{\PRD}{{\em Phys. Rev.} D}
\newcommand{\ZPC}{{\em Z. Phys.} C}

\end{document}